\documentclass{article}
\usepackage{hiph-preprint}
\volnumber{22} \issuenumber{1} \edyear{2005}                             
\frompage{000} \topage{000}                                              
\recrevdate{1 January 2005}                                              
\def\rightmark{Transport theory from multiple scattering expansion}
\def\leftmark{J.-S. Gagnon, F. Fillion-Gourdeau and S. Jeon}
 
\title{All orders transport theory from the multiple scattering expansion of the self-energy: the central cuts} 
\authors{ 
{Jean-S\'{e}bastien Gagnon$^1$, Fran\c{c}ois Fillion-Gourdeau$^1$ and Sangyong Jeon$^{1,2}$  %
\index{One, J.-S. Gagnon} 
\index{Two, F. Fillion-Gourdeau} 
\index{Three, S. Jeon}
}\\[2.812mm]
{\normalsize
\hspace*{-8pt}$^1$ Department of Physics, McGill University,\\ 
3600 University street Montr\'{e}al, Canada\\[0.2ex] 
\hspace*{-8pt}$^2$ RIKEN-BNL Research Center,\\ 
11973-5000 Upton, U.S.A.
}}
 
\abstract{We use the full multiple scattering expansion of the retarded self-energy to obtain the gain and loss rates present in the Kadanoff-Baym relativistic transport equation.  The rates we obtain include processes with any number of particles.  As a first approximation, we only consider central cuts in the self-energies, but otherwise our results are general.  We specialize to the case of scalar field theory to compare with lowest order results.  The main application of this work is relativistic transport theory of very dense systems, such as the quark-gluon plasma or the early universe, where multi-particle interactions are important.}

 
\makeindex
\begin{document}
 
\maketitle

\section{Introduction}\label{intro}

Transport theory is an important tool in the study of the quark-gluon plasma (e.g. \cite{Geiger_1995,Molnar_Gyulassy_2000}).  In the case of the quark-gluon plasma (QGP), it is used to model some aspects of its evolution.  In particular, in QGP thermalization studies, many authors (e.g. \cite{Xu_Greiner_2005,AMY_2003_2,Baier_etal_2001}) stress the importance of including multi-particle interactions such as $gg \rightarrow ggg$ in transport theory models.  This is to be expected, since at the beginning of the QGP expansion, its energy density is $\geq 1$ GeV/fm$^{3}$ (approximately 6 times the density of cold nuclear matter).  The same conclusion is drawn in transport coefficients computations \cite{Jeon_Yaffe_1996,AMY_2003_2}.  Thus including multi-particle interactions in kinetic equations seems to be an important ingredient in the study of very dense systems.

The goal of this paper is to show how to obtain complete relativistic gain/loss rates containing any number of particles.  These rates are the ones present in the collision term of a relativistic Kadanoff-Baym transport equation.  Our derivation is based on the exact multiple scattering expansion of the self-energy \cite{Jeon_Ellis_1998}.

 \section{Relativistic Transport Equations}
\label{Transport_equations}

The results presented in this section rely on the work of \cite{Carrington_etal_2005}.  In this paper, the authors derive a relativistic transport equations from quantum field theory for a system of real scalar fields (cubic and quartic interactions):
\begin{eqnarray}
\label{Baym_Kadanoff}
E_{p}\left( \frac{\partial}{\partial t}+ \mathbf{v} \cdot \nabla \right) f(X,\mathbf{p}) +\nabla V(X,\mathbf{p}) \nabla_{p} f(X,\mathbf{p})
& = & \theta(p_{0}) C(X,p)
\end{eqnarray}

\noindent along with the mass-shell condition $p^{2}-m^{2}+V(X,{\mathbf p}) = 0$, where $E_{p}$ is the solution of the mass-shell equation, $\mathbf v \equiv \partial E_{p}/\partial {\mathbf p}$ is the velocity, $f(X,{\mathbf p})$ is the distribution function, $V(X,{\mathbf p})$ is the Vlasov term (corresponding to an effective mass) and $C(X,p)$ is the collision term (given below).  To obtain Eq. \ref{Baym_Kadanoff} and the mass-shell condtion, one must assume weak inhomogeneity with respect to the inverse characteristic momentum (allows for the gradient expansion) and the Compton wavelength (justifies the quasi-particle approximation) and also weak coupling.  The collision term in Eq. \ref{Baym_Kadanoff} is given by:
\begin{eqnarray}
\label{collision}
C(X,p) & = & -\frac{i}{2} \left[\Pi^{<}(X,p) \left(f(X,\mathbf{p})+1\right) - \Pi^{>}(X,p) f(X,\mathbf{p})\right]
\end{eqnarray}

\noindent where $\Pi^{<,>}(X,p)$ are Wightman functions.  These Wightman functions are ``completely cut'' self-energies and are interpreted as gain/loss rates.  These microscopic collision rates must be computed from quantum field theory.  In \cite{Carrington_etal_2005}, the authors use perturbation theory and compute the rates explicitly.  In this paper, we instead use the full Multiple Scattering Expansion of the Self-Energy \cite{Jeon_Ellis_1998} to derive the rates and express them in terms of physical processes.

\section{The Multiple Scattering Expansion of the Self-Energy}
\label{Multiple_scattering}

The idea behind the expansion is simple.  For simplicity, let's consider only scalar theories.  The starting point is the common structure of the four bosonic propagators corresponding to the contour Green's function in the real time formalism.  Each propagator can be decomposed into a zero and finite temperature part, i.e. $G^{ij}(p) = G_{0}^{ij}(p) + \Gamma(p)$, the latter being common to the four propagators.  Using this property, it is possible to reorganize any self-energy diagram as a power series in the number of thermal phase space factor $\Gamma(p) = n(|p^{0}|)2\pi\delta(p^{2}-m^{2})$, where the coefficients of the expansion are zero temperature scattering diagrams \cite{Jeon_Ellis_1998}.

\begin{figure}[htb]
\epsfxsize=10cm
\centerline{\epsfbox{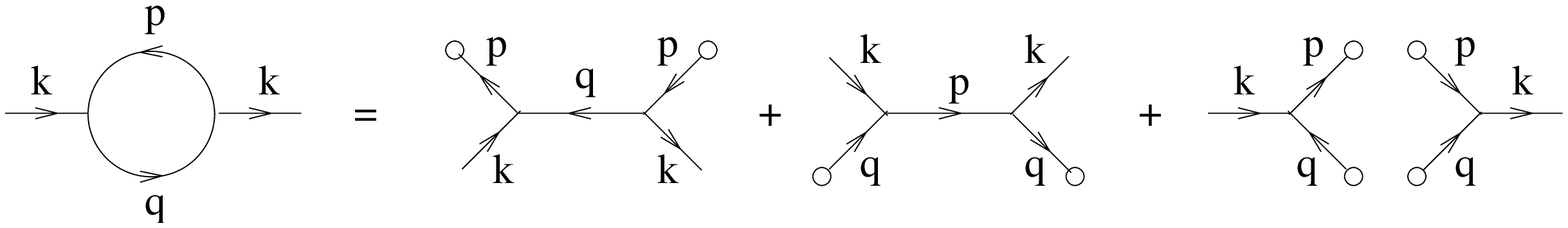}}
\caption[]{One-loop self-enegy example of multiple scattering expansion.  The white circles correspond to thermal lines.  The first and second terms are first order in the thermal expansion, while the third term is second order (the zero temperature term is not shown)}
\label{One_loop_expansion}
\end{figure}
To compute the rates $\Pi^{<,>}(k)$ appearing in Eq. \ref{collision}, we need the multiple scattering expansion of the imaginary part of the retarded self-energy.  Following the procedure in \cite{Jeon_Ellis_1998} and concentrating on the loss rate, the thermal contribution to $\Pi^{>}(k)$ can be written as (a similar expression exists for $\Pi^{<}(k)$ \cite{Jeon_Ellis_1998}):
\begin{eqnarray}
\label{Scattering_expansion}
\Pi^{>}(k) & = & \frac{1}{S_{\{l_{i}\}}} \sum_{n=0}^{\infty} \left(\prod_{i=1}^{n+1}\int d\Gamma_{i}\right) \langle k,\{l_{i}\}|{\cal T}_{k}^{\dagger}{\cal T}_{k}|k,\{l_{i}\} \rangle_{\rm conn} 
\end{eqnarray}

\noindent where $k$ is the external momentum, $l_{i}$'s are momenta corresponding to thermal lines, $S_{\{l_{i}\}}$ are symmetry factors, $d\Gamma_{i}^{\sigma} = \frac{d^{3}l_{i}}{(2\pi)^{3}2E_{i}}n(E_{i})$, $E_{i} = \sqrt{{\bf l}_{i}^{2}+m_{i}}$, ${\cal T}_{k}$ is the part of the ${\cal T}$ matrix involving $k$ (when it is in the initial state) and $\sum_{n=0}^{\infty}$ is the summation over the expansion parameter $d\Gamma_{i}$.  The subscript ``conn'' means fully connected diagrams only.  The operators inside the bra-kets are related to the usual ${\cal T}$ matrix containing {\em all} interactions.



\section{Gain/Loss Rates}
\label{Rates}

Let us consider the loss rate.  The idea is to start from Eq. \ref{Scattering_expansion} and insert a complete set of many-particle states between the two ${\cal T}_{k}$ operators.  This gives:
\begin{eqnarray}
\label{Loss_term_1}
\Pi^{>}(k) & = & \sum_{n=0}^{\infty} \sum_{s=0}^{\infty} \frac{1}{n!} \left(\prod_{i=1}^{n}\int\frac{d^{3}l_{i}}{(2\pi)^{3}}\frac{n(E_{i})}{2E_{i}}\right) \left(\frac{1}{s!}\prod_{j=1}^{s}\int\frac{d^{3}p_{j}}{(2\pi)^{3}}\frac{1}{2E'_{j}}\right)  \nonumber \\
	   &    & \times \langle k,\{l_{i}\}|{\cal T}_{k}^{\dagger}|\{p_{j}\}\rangle \langle \{p_{j}\}|{\cal T}_{k}|k,\{l_{i}\} \rangle_{\rm conn}
\end{eqnarray}

\noindent where the $1/n!$ comes from the symmetry factor of the cut diagrams \cite{Jeon_Ellis_1998}.  It is understood that a product without a factor is one.  This operation has the effect of ``cutting'' the diagrams on the RHS of Fig. \ref{One_loop_expansion}.  Let us concentrate on ``central cuts'' only, i.e. diagrams where only the lines that are already cut can be replaced by $\Gamma(p)$; we treat the case of non-central cuts in another paper \cite{Fillion_etal_2005}.  For the central cuts case, $n_{f}$ number of $l_{i}$'s pass through ${\cal T}_{k}$ (they are spectators) and replace $n_{f}$ of the $p_{j}$'s.  Inserting the factor $1 = \sum_{n_{i}}^{\infty} \delta_{n-n_{f}-n_{i}}\theta(n \geq n_{f})$ to reorganize the sum and after a bit of algebra, we get:
\begin{eqnarray}
\label{Loss_term_2}
\Pi_{\rm central}^{>}(k) & = & \sum_{s=0}^{\infty} \sum_{n_{i}=0}^{\infty} \frac{1}{n_{i}!s!} \left(\prod_{i=1}^{n_{i}}\int\frac{d^{3}l_{i}}{(2\pi)^{3}}\frac{n(E_{i})}{2E_{i}}\right) \left(\prod_{j=1}^{s}\int\frac{d^{3}p_{j}}{(2\pi)^{3}}\frac{(1+n(E'_{j}))}{2E'_{j}}\right)  \nonumber \\
	   &    & \times \langle k,\{l_{i}\}|{\cal M}_{k}^{\dagger}|\{p_{j}\}\rangle \langle \{p_{j}\}|{\cal M}_{k}|k,\{l_{i}\} \rangle_{\rm conn}
\end{eqnarray}

\noindent where the ${\cal T}_{k}$ matrix elements become ${\cal M}_{k}$ after all spectators are passed through (the $\delta$ function associated with ${\cal M}_{k}$ is not written explicitly).  Equation \ref{Loss_term_2} is expressed in terms of physical scattering processes and contains any number of particles (the indices $s$ and $n_{i}$ specify the number of final and initial particles, respectively).  A similar expression can be obtained for the gain term \cite{Fillion_etal_2005}.  We can see that it reproduces the two-loop and three-loop results of \cite{Carrington_etal_2005} for $n_{i}=1,\; s=2$ and $n_{i}=1,\; s=3$.

\section{Conclusion}
\label{Conclusion}

We have shown that the multiple scattering expansion of the retarded self-energy can be used to obtain the gain/loss rates of the relativistic Kadanof{}f-Baym equation.  These rates have a direct interpretation in terms of physical processes.  Our approach has several advantages: the calculation is completely general, it contains any number of particles, it is valid to all orders and it is in principle generalizable to gauge theories.  However, there is an important caveat: we assume that pinch singularities vanish, even if the system is out-of-equilibrium.  This assumption is not true in general and we are currently working on improving this aspect of the problem.  Details of the derivation and the generalization to non-central cuts will be discussed in another paper \cite{Fillion_etal_2005}.

\section*{Acknowledgment(s)}

The authors are supported in part by the Natural Sciences and Engineering Research Council of Canada and by le Fonds Nature et Technologies du Qu\'ebec. S.J. also
thanks RIKEN BNL Center and U.S. Department of Energy [DE-AC02-98CH10886] for
providing facilities essential for the completion of this work.

\vfill\eject
\end{document}